\documentclass[%
 reprint,
https://www.overleaf.com/project/5f7ddcf04978e9000171c65c%preprint,
 amsmath,amssymb,
 aps,
]{revtex4-2}

\usepackage{graphicx}% Include figure files
\usepackage{dcolumn}% Align table columns on decimal point
\usepackage{bm}% bold math
\newcommand{\pt}{p_{\mathrm{T}}}
\usepackage[mathlines]{lineno}% Enable numbering of text and display math
\begin{document}

\preprint{APS/123-QED}

\title{Unsupervised clustering for collider physics}% Force line breaks with \\

\author{Vinicius Mikuni}
 \email{vinicius.massami.mikuni@cern.ch}
\author{Florencia Canelli}%
\affiliation{%
 University of Zurich,\\Winterthurerstrasse 190, 8057 Zurich, Switzerland
}%

\date{\today}% It is always \today, today,

\begin{abstract}
We propose a new method for Unsupervised Clustering for collider  physics named UCluster, where information in the embedding space created by a neural network is used to categorize collision events into different clusters that share similar properties. We show how this method can be developed into an unsupervised multiclass classification of different processes and applied in anomaly detection of events to search for new physics phenomena at colliders. 
\end{abstract}

\maketitle

%\tableofcontents

\section{Introduction}
\label{sec:intro}
The Standard Model (SM) of particle physics has been successful so far at describing the interaction of fundamental particles in high energy physics (HEP). The ATLAS \cite{Aad:2008zzm} and CMS \cite{Chatrchyan:2008aa} Collaborations have tested the SM extensively using particle collision events at the CERN Large Hadron Collider (LHC), while also looking for deviations from the SM that could point to physics beyond the SM (BSM). Since the underlying nature of the new physics is not known, new methods designed to be model independent have proliferated in the recent years. These strategies aim at finding deviations or detecting anomalies where only SM events are used and avoiding any dependence on BSM signals. For a short review of recent approaches see \cite{PhysRevD.101.075042}. 
% mostly well understood?

For measurements of SM parameters, a fully unsupervised multiclass classification method would be advantageous. This is particularly true for precision measurements of SM parameters. Simulations are often needed to describe the properties of different processes produced in the LHC collisions. However, simulated events are not always precise in all physics process. This can be caused either by a lack of simulated events compared to the data expectation, or the need of corrections that are beyond the accuracy of the approximations used in the simulation. Further precision might be 
computationally prohibitive to achieve, or beyond the capability of our current methods. To mitigate these issues, different data-driven methods often replace the event simulations. See \cite{Aaboud:2017hdf,Sirunyan:2019jud,Sirunyan:2019vgj,Aad:2020cws} for recent examples. 

When two or more processes are not well modeled, the common approach is to design multiple control regions, often defined using high level distributions, to create a high purity sample that allows a data driven estimation and modeling for this process. However, since it is not always straightforward to define each of these regions without relying on simulations, an unsupervised multiclass classification approach could be used instead.

In this paper, we introduce a method for unsupervised clustering (UCluster). The main idea of UCluster is to use a neural network (NN) to reduce the data dimensionality  while retaining the main properties of the data set. In this reduced representation, a clustering objective is added to the training to encourage points embedded in this space to be close together when they share similar properties and far apart otherwise. We test the performance of UCluster in the context of two different tasks: unsupervised multiclass classification of three different SM processes and unsupervised anomaly detection.

\section{Related works}
\label{sec:related}
Recently, different and innovative strategies have been proposed for unsupervised training in HEP, mostly in the context of event classification. A few examples of methods exploiting anomaly detection signatures as over-densities are \cite{Metodiev:2017vrx, Collins:2019jip} and more recently \cite{PhysRevD.101.075042}. In these approaches, anomalous events are identified as localized excesses in some distribution, where machine learning is then used to enhance the local significance of the new physics process.

While many strategies focus on unsupervised anomaly detection, other methods have also been proposed to better understand SM processes without relying on simulation, like the work developed in \cite{Metodiev_2018} for quark and gluon classification with jet topics and the methods developed in \cite{PhysRevD.100.056002}, employing latent Dirichlet allocation to build a data-driven top-quark tagger. In order to create an unsupervised and model independent approach, the majority of the strategies rely on binary classification, where the main goal is to test if an event (or a group of events) resulting from a particle collision is compatible with one out of two competing hypotheses. Approaches applied to mixed samples with more than two components were also studied in \cite{JMLR:v10:quadrianto09a, NIPS2014_5453}, where prior knowledge of the label proportion for each component in the mixed sample is required to achieve a good performance. 

In this work, we propose an unsupervised method for multiclass classification whose only requirement is on the expected number of different components inside a mixed sample. The same method is applied to anomalous event detection, where the data is partitioned into clusters that isolate the anomaly from backgrounds.

\section{Method description}
\label{sec:method}
UCluster consists of two components: a classification step to ensure events with similar properties are close in the embedding space created by a NN; and a clustering step, where the network learns to cluster embedded events of similar properties. These two tasks are accomplished by means of a combined loss function containing independent components to guarantee each of the described steps.

The classification loss ($L_{\mathrm{focal}}$), applied to the output nodes of a NN, is defined by the focal loss \cite{DBLP:journals/corr/abs-1708-02002}. The focal loss improves the classification performance for unbalanced labels, the case for the classification tasks to be introduced in the following sections. The expression for the focal loss is:

\begin{eqnarray}
     L_{\mathrm{focal}} &&=  -\frac{1}{N}\sum_j^N \sum_m^M  y_{j,m} (1 - p_{\theta,m}(x_j))^{\gamma} \nonumber\\
    && \times  \log(p_{\theta,m}(x_j)),
    \label{eq:class_loss}
\end{eqnarray}

where $p_{\theta,m}(x_j)$ is the network’s confidence, for event $x_j$ with trainable parameters $\theta$, to be classified as class $m$. The term $y_{j,m}$ is 1 if class $m$ is the correct assignment for event $j$ and 0 otherwise.
In this work, we fix the hyperparameter $\gamma=2$ of the focal loss.

The clustering loss ($L_{\mathrm{cluster}}$) is defined similarly as the loss developed in \cite{DBLP:journals/corr/abs-1806-10069}:
\begin{equation}
    L_{\mathrm{cluster}} = \frac{1}{N}\sum_k ^K \sum_j^n \left \| f_{\theta}(x_j) - \mu_k \right \|^{2}\pi_{jk},
    \label{eq:cluster_loss}
\end{equation}
where the distance between each event $j$ and each cluster centroid $\mu_k$ is calculated in the embedding space $f_{\theta}$ of the neural network with trainable parameters $\theta$. 
The function $\pi_{jk}$ weighs the importance of each event and takes the form:
\begin{equation}
    \pi_{jk} = \frac{e^{-\alpha \left \| f_{\theta}(x_j) - \mu_k \right \|}}{\sum_{k'} e^{-\alpha \left \| f_{\theta}(x_j) - \mu_k \right \|}},
\end{equation}
with hyperparameter $\alpha$ identified as an inverse temperature term. Since $L_{\mathrm{cluster}}$ is differentiable, stochastic gradient descent can be used to optimize jointly the trainable parameters $\theta$ and the centroid positions $\mu_k$. 

The combined loss to be minimised is:
\begin{equation}
    L =  L_{\mathrm{focal}} + \beta L_{\mathrm{cluster}}.
    \label{eq:loss}
\end{equation}
The hyperparameter $\beta$ controls the relative importance between the two losses. For these studies, we fix $\beta$=10 to ensure that both components have the same order of magnitude.

Since $L_{\mathrm{cluster}}$ requires an initial guess for the centroid positions, we pre-train the model using only $L_{\mathrm{focal}}$ for 10 epochs. After the pre-training, the K-Means algorithm \cite{10.2307/2346830} is applied to the object embeddings to initialize the cluster centroids. The full training is then carried out with the combined loss defined in Eq.~\ref{eq:loss}. To allow the cluster centers to change, the inverse temperature $\alpha$ has a starting value of 1 and linearly increases by 2 for each following epoch. 

\begin{figure*}[htb]
    \centering
    \includegraphics[width=0.8\textwidth]{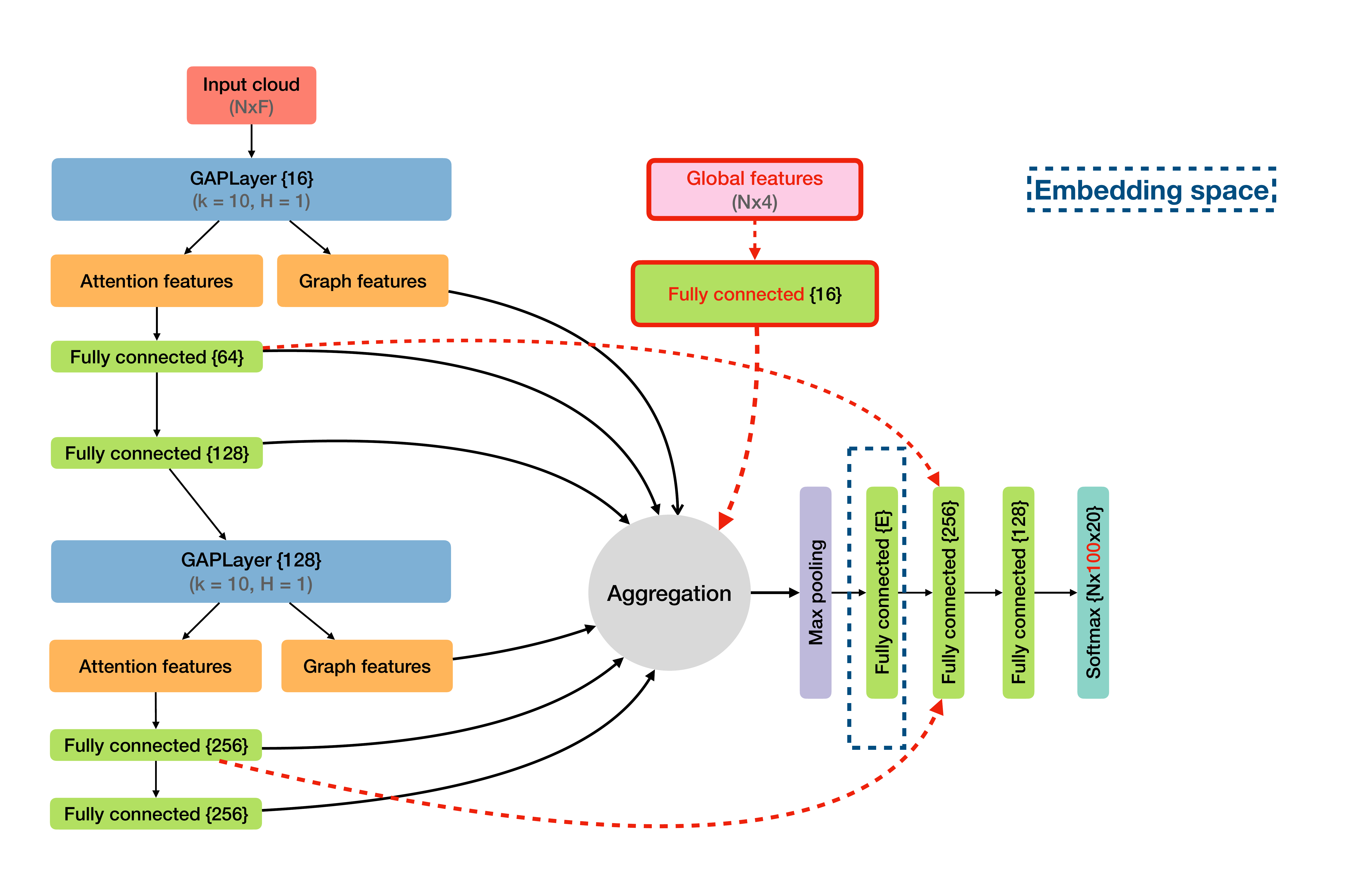}
    \caption{ABCNet architecture used in UCluster for a batch size N, F input features, and embedding space of size E. Fully connected layers and encoding node sizes are denoted inside ``\{\}''. For each GAPLayer, the number of k-nearest neighbors (k) and heads (H) are given. The additional components used only for anomaly detection are shown in red.}
    \label{fig:abc_model}
\end{figure*}

\section{General implementation}
The implementation of UCluster is done using ABCNet \cite{Mikuni:2020wpr}. ABCNet is a graph-based neural network where each reconstructed particle is taken as a node in an graph. The importance of each node is then learned by the model by the usage of attention mechanisms.  The embedding space for the clustering loss in Eq.~\ref{eq:cluster_loss} is taken as the output of a max-pooling layer. For the following studies, the 10 nearest neighbours from each particle are used to calculate the  GAPLayers \cite{2019arXiv190508705C}. The initial distances are calculated in the pseudorapidity-azimuth ($\eta-\phi$) space using the distance $\Delta R = \sqrt{\Delta\eta^2 + \Delta\phi^2}$. The second GAPLayer uses the Euclidean distances in the space created by subsequent fully connected layers. The architectures used for multiclass classification and anomaly detection are depicted in Fig.~\ref{fig:abc_model}. Besides the output classification size, both tasks share almost identical architectures. The model used for anomaly detection uses additional high-level distributions and additional skip connections after the pooling layer to improve the classification performance. In both cases the batch size is set to 1024 and the training is stopped after for 100 epochs.

ABCNet is implemented in Tensorflow v1.14 \cite{tensorflow2015-whitepaper}. An Nvidia GTX 1080 Ti graphics card is used for the training and evaluation steps. For all tasks described in this paper, the Adam optimizer \cite{2014arXiv1412.6980K} is used. The learning rate starts from 0.001 and decreases by a factor 2 every three epochs, until reaching a minimum of 1e-5.

\section{Unsupervised multiclass classification}
\label{sec:multiclass}
\begin{figure}[ht]
    \centering
    \includegraphics[width=0.45\textwidth]{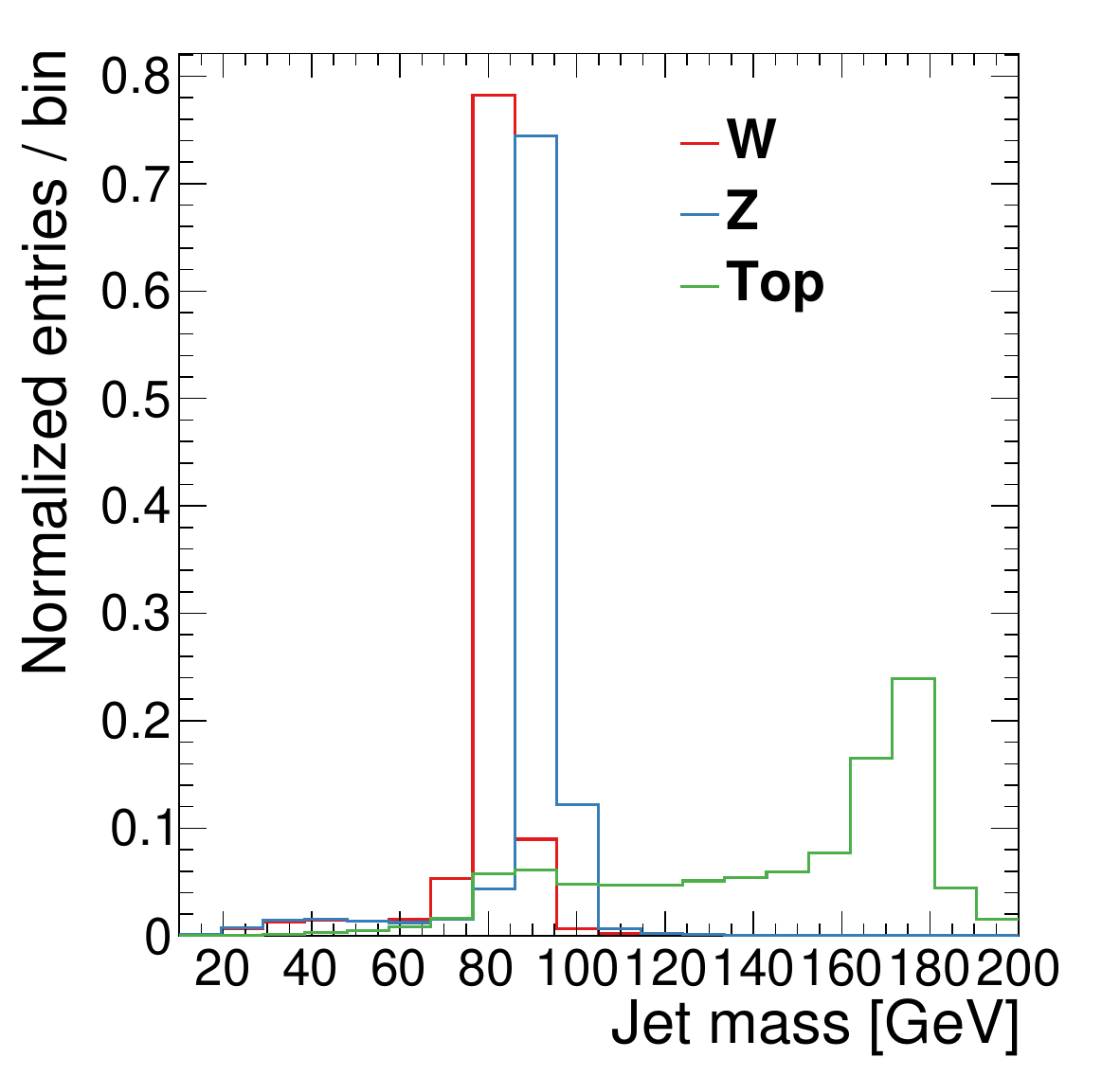}
    \caption{Normalized distribution of the jet mass  of each category used in the unsupervised multiclass classification task. The bin boundaries represent the boundaries used to define the jet mass labels.}
    \label{fig:jet_mass}
\end{figure}

The applicability of UCluster is demonstrated on an important problem in high energy physics: unsupervised multiclass classification. To achieve good performance, we require a task that results in a suitable embedding space. This task should be such that events stemming from the same physics process are found close together in the embedding space as compared to events from different physics processes. Here, a jet mass classification task is chosen in order to provide meaningful event embeddings. Given a set of particles belonging to a jet, we ask our model to correctly identify the invariant mass of the jet. This task chosen is inspired by the correlation of jet substructure observables and the invariant mass of a jet \cite{Dolen:2016kst,Komiske:2019fks}. The goal is to have our machine learning method learn to extract relevant information regarding the different jet substructures by first learning how to correctly identify the mass of a jet. The simplest solution to this problem could be achieved by the four-vector sum of all the particle's constituents, leading to an embedding space that does not have separation power for different types of jets. To alleviate this issue, we instead define a jet mass label by taking 20 equidistant steps from 10 to 200 GeV, as shown in Fig.~\ref{fig:jet_mass}.  The task is then to identify the correct mass interval a jet belongs to, instead of the specific mass value. The input distributions used for the training are listed in Table~\ref{tab:multi_vars}.

\begin{table*}[ht]
    \centering
    \caption{Description of each feature used to define a point in the ABCNet implementation for unsupervised multiclass classification.}
    \label{tab:multi_vars}
	\begin{tabular}{lc}
    \hline
             Variable & Description  \\
            \hline
            $\Delta\eta$       &  Difference between the pseudo-rapidity of the constituent and the  jet\\  
            $\Delta\phi$       &  Difference between the azimuthal angle of the constituent and the  jet\\  
            $\log\pt$       &  Logarithm of the constituent's $\pt$ \\  
            $\log\mathrm{E}$       &  Logarithm of the constituent's E \\  
            $\log\frac{\pt}{\pt(\mathrm{jet})}$       &  Logarithm of the ratio between the constituent's $\pt$ and the  jet $\pt$\\  
            $\log\frac{\mathrm{E}}{\mathrm{E}(\mathrm{jet})}$       &  Logarithm of the ratio between the constituent's E and the jet E\\  
            $\Delta\mathrm{R}$       & Distance in the $\eta-\phi$ space between the constituent and the  jet\\  
             PID       &  Particle type identifier as described in \cite{Tanabashi:2018oca}. \\

	\end{tabular}
\end{table*}

For this study, a sample containing simulated jets originating from W bosons, Z bosons, and top quarks produced at $\sqrt{s}$ = 13 TeV proton-proton collisions is used. This data set is created and configured using a parametric description of a generic LHC detector, described in \cite{Coleman:2017fiq,Duarte:2018ite}. The jets are clustered with the anti-kt algorithm \cite{Cacciari:2008gp} with radius parameter R = 0.8, while also requiring that the jet's $\pt$ is around 1 TeV, ensuring that most of the decay products of the generated particles are found inside a single jet. 

The samples are available at \cite{pierini_maurizio_2020_3602254}. For each jet, up to 100 particles are stored. If more particles were found inside a jet, the event is truncated, otherwise zero-padded up to 100. The training set contains 300,000 jets, while the validation sample consists of 140,000 jets.

\begin{figure*}[ht]
    \centering
    \includegraphics[width=0.4\textwidth]{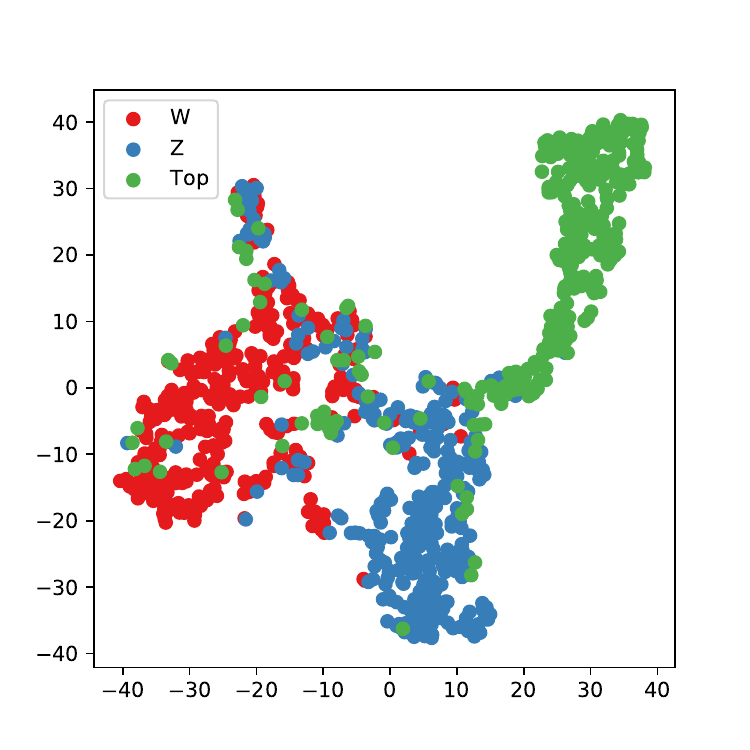}
    \includegraphics[width=0.4\textwidth]{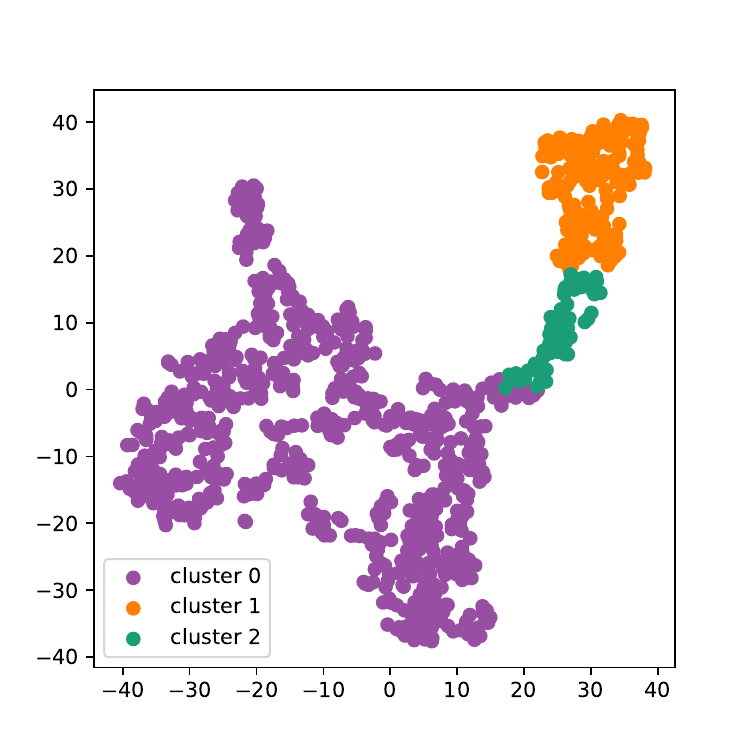}
    \includegraphics[width=0.4\textwidth]{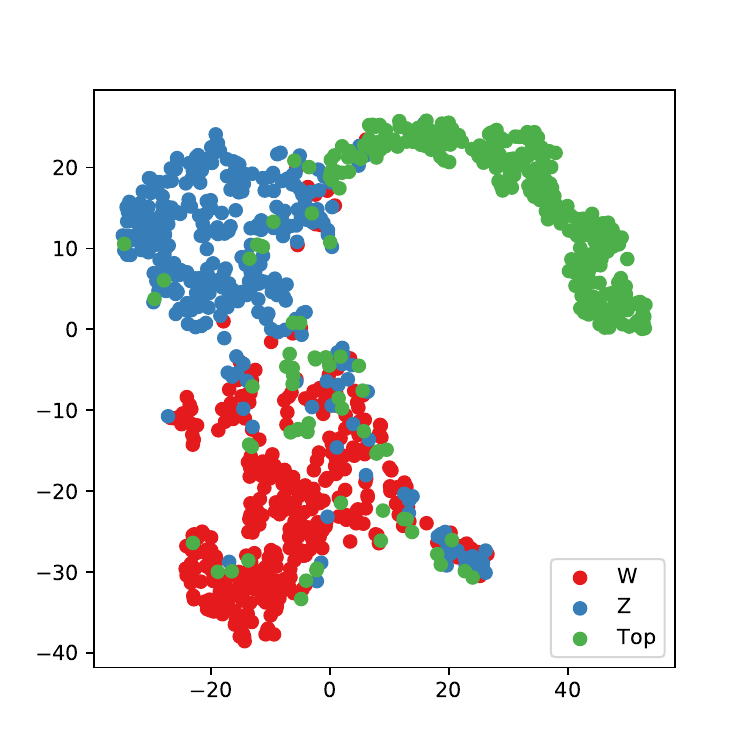}
    \includegraphics[width=0.4\textwidth]{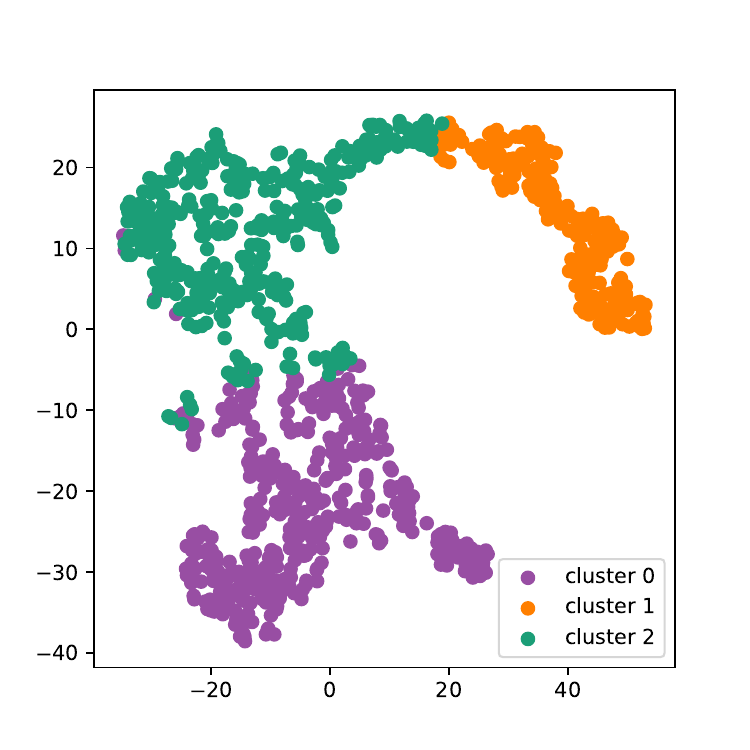}
    \caption{t-SNE visualization of the embedding space after the pre-training and before the full training (top row) and after the full training (bottom row) for multiclass classification with 1000 jets. The true label information is shown on the left, while the initial cluster labels using a k-Means approach is shown on the right.}
    \label{fig:tsne_multi}
\end{figure*}

To visualize the embedding space, the t-SNE visualization method \cite{vanDerMaaten2008} is used for 1000 jets, taken just after the pre-training with only the classification loss, and compared to the space created after full training is performed. After the pre-training, the initial label assignment is taken from a k-Means approach, shown in Fig.~\ref{fig:tsne_multi} (top right) while the true labels are shown in Fig.~\ref{fig:tsne_multi} (top left). At this stage, the clustering accuracy, calculated using the Hungarian algorithm \cite{doi:10.1002/nav.3800020109}, is 51\%. After the full training is performed, the trained labels are shown in Fig.~\ref{fig:tsne_multi} (bottom right) with a clustering accuracy of 81\% compared to the true label assignment in Fig.~\ref{fig:tsne_multi} (bottom left). 

To inspect the quality of the embedding space further, a supervised KNN is trained using only the embedding features as inputs. Its performance is then compared to a separate KNN with the same setup, but using only the jet mass as input. The supervised KNNs are trained to determine class membership given the label of the 30 nearest neighbors. For the training, 35k events are used and tested on an independent sample with 15k events.

\begin{figure*}[ht]
    \centering
    \includegraphics[width=0.4\textwidth]{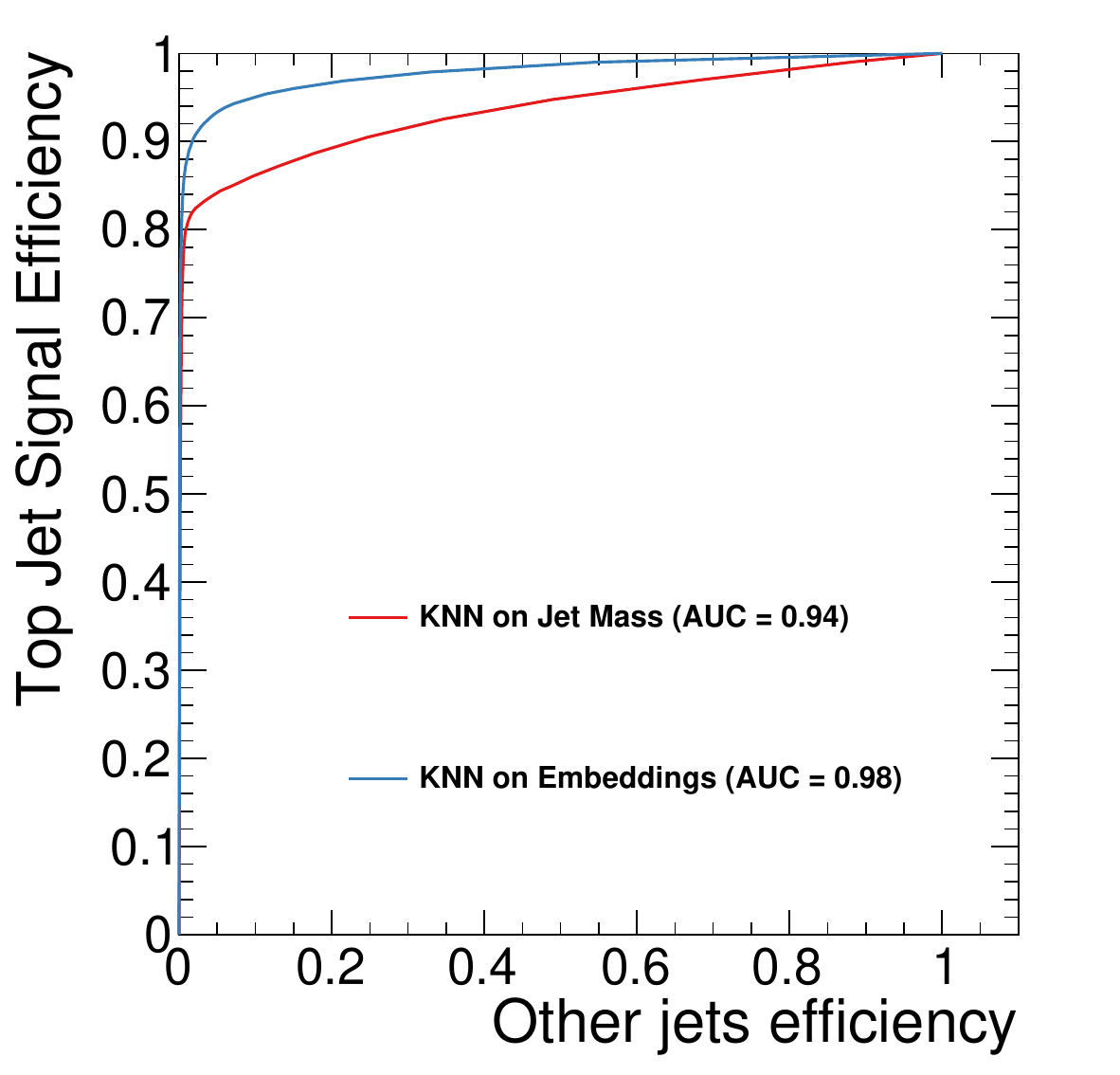}
    \includegraphics[width=0.4\textwidth]{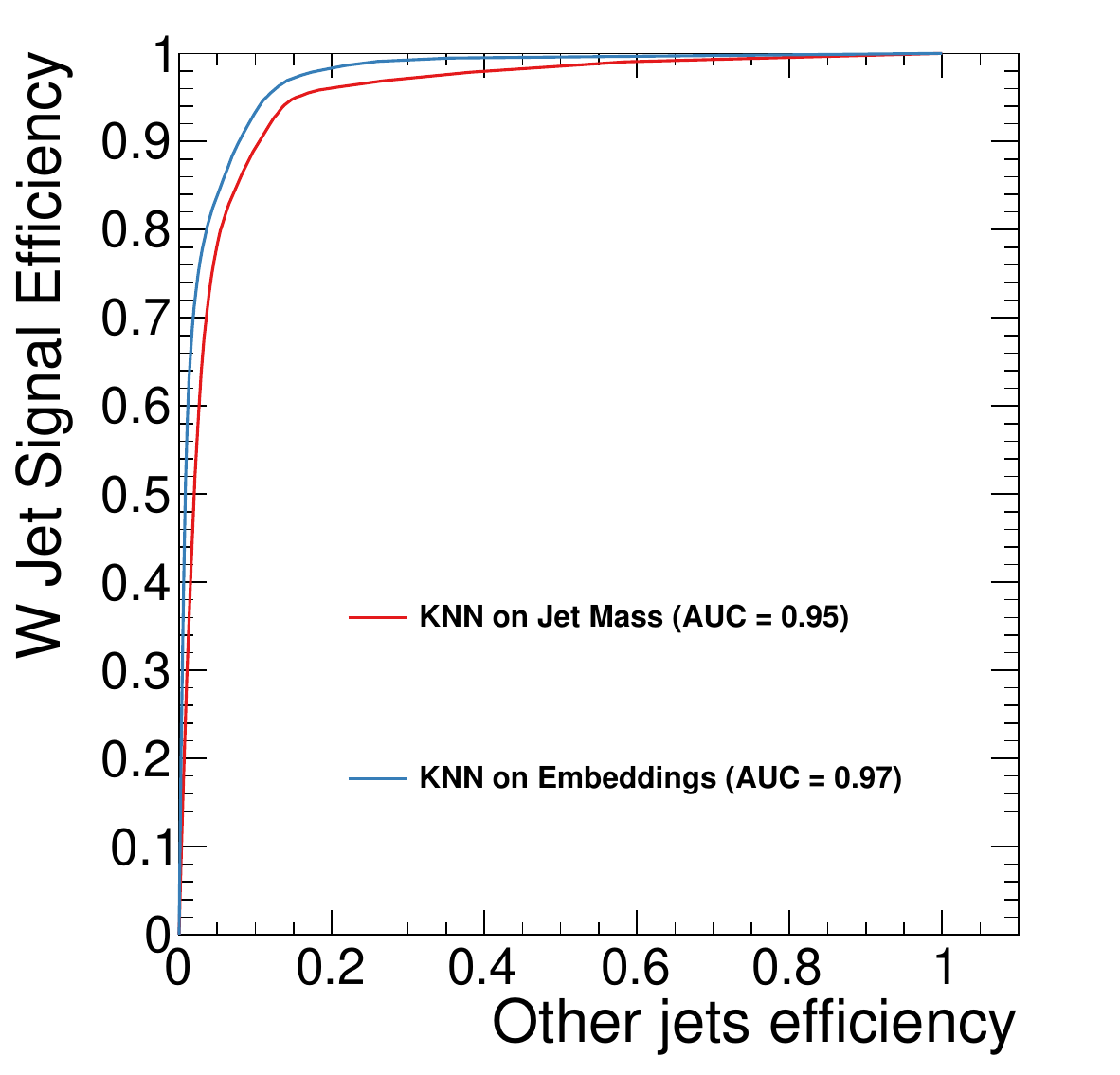}
    \includegraphics[width=0.4\textwidth]{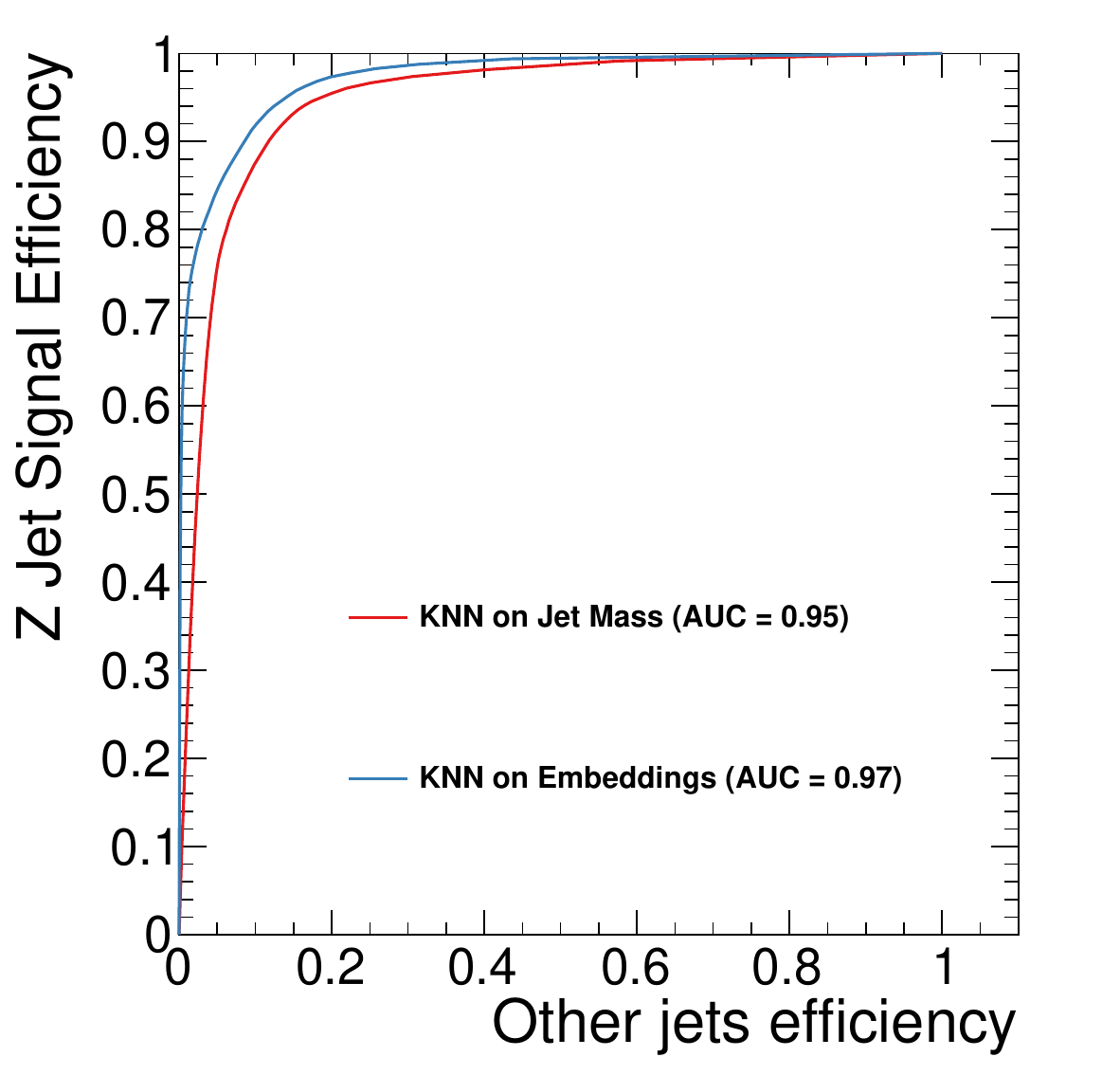}
    \caption{ROC curves for each jet category when considering the other jet categories as a background.}
    \label{fig:roc}
\end{figure*}

The one-vs-all performance is compared using a receiver operating characteristic (ROC) curve in Fig.~\ref{fig:roc}, where one category is considered the signal of interest while the others are considered a background. The area under curve (AUC) for each process is also shown. The resulting AUC for the supervised training using the event embeddings is higher than the jet mass alone for all categories. Top quark classification shows a  particularly large improvement by using the embedding space information. We attribute this improvement to jets containing a top quark showing a broader mass distribution compared to W and Z bosons, resulting in a worse invariant mass separation  as seen in Fig.~\ref{fig:jet_mass}. UCluster is able to learn other jet properties beyond the invariant mass, improving the overall performance.

To estimate an upper bound on the UCluster performance, a fully supervised model using the full ABCNet architecture is also trained. The ABCNet architecture is used to train a classifier containing the real class labels as targets, achieving an accuracy of 92\%. The comparable results between the fully supervised approach and the KNN trained on the event embeddings demonstrate how the method is able to reduce the dimensionality of the input data while retaining relevant information.

The accuracies achieved with the full supervision and the other approaches are summarized in Tab.~\ref{tab:multi_acc}.

\begin{table}[ht]
    \centering
    \caption{Supervised and unsupervised clustering accuracy of UCluster when using only the embedding space features.}
    \label{tab:multi_acc}
	\begin{tabular}{lc}
    \hline
            Algorithm  & Accuracy \\
            \hline
            Pre-training k-Means & 51\% \\
            UCluster & 81\% \\
            Supervised KNN & 89\% \\
            Supervised training & 92\% \\

	\end{tabular}
\end{table}

\section{Anomaly detection}
\label{sec:anomaly}
UCluster can also be applied to anomaly detection. Here, we show an example where anomalous events, created from an unknown physics process, are found to be close in the embedding space created from a suitable classification task. This technique is motivated by the fact that, irrespective to the underlying physics model, events created by the same physics process carry similar event signatures. 

To create a suitable embedding space we modify the approach described in Sec.~\ref{sec:multiclass} to take into account all the particles created in a collision event rather than a single jet. To do so, the classification task is instead changed to a part segmentation task. We consider all particles associated to a clustered jet. Each particle then receives a label proportional to the mass of the jet that it was clustered into. For this task, we require the model to learn not only the mass of the associated jet the particle belongs to, but to also to learn which particles should belong to the same jet. This approach is motivated by the fact that jet substructure often contains useful information for distinguishing different physics processes, as studied in the previous section.

\begin{figure}[ht]
    \centering
    \includegraphics[width=0.45\textwidth]{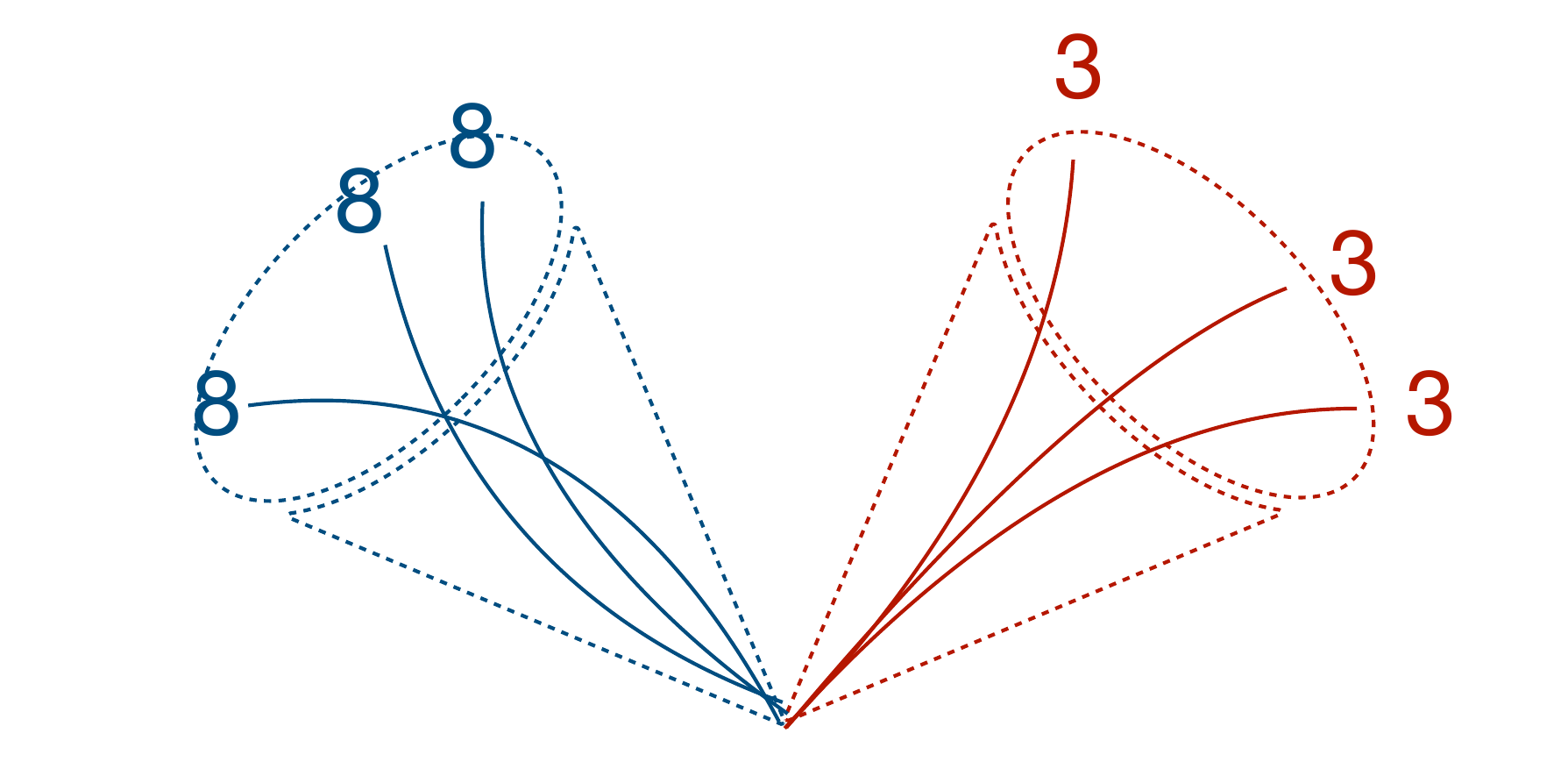}
    \caption{Schematic of the labels for anomaly detection. Each particle associated to a clustered jet receives a mass label proportional to the respective jet mass. The larger the number, the more massive the associated jet.}
    \label{fig:part_label}
\end{figure}

The mass labels are then created by defining 20 equidistant intervals from 10 to 1000 GeV. For simplicity, only the two heaviest jets are considered per event. A simplified example of the label definition is shown in Fig.~ \ref{fig:part_label}.

To perform these studies, we use the R\&D data set created for the LHC Olympics 2020 \cite{gregor_kasieczka_2019_2629073}. The data set consists of a million quantum chromodynamic (QCD) dijet events simulated with Pythia 8\cite{Sjostrand:2014zea} without pile up or multiple parton interactions. The BSM signal consists of a  hypothetical W' boson with mass $m_W$ = 3.5 TeV that decays into an X and Y bosons with masses $m_X$ = 500 GeV and  $m_Y$ = 100 GeV, respectively. The X and Y bosons, on the other hand, decay promptly into quarks. The detector simulation is performed with Delphes 3.4.1 \cite{deFavereau:2013fsa} and particle flow objects are clustered into jets using the Fastjet \cite{Cacciari:2011ma} implementation of the anti-kt algorithm with R = 1.0 for the jet radius. Events are required to have at least one jet with $\pt$>1.3 TeV. The number of signal events generated is set to as 1\% of the total number of events. From this data-set, 300k events are randomly selected for training, 150k for testing and 300k events, are used to evaluate the clustering performance.

The distributions used as an input for ABCNet are described in Tab.~\ref{tab:anomaly_vars}. To improve the clustering performance, a set of high level variables are added to the network. The goal of the additional distributions is to parameterize the model performance as described in \cite{Baldi:2016fzo}.

\begin{figure*}[t]
    \centering
    \includegraphics[width=0.40\textwidth]{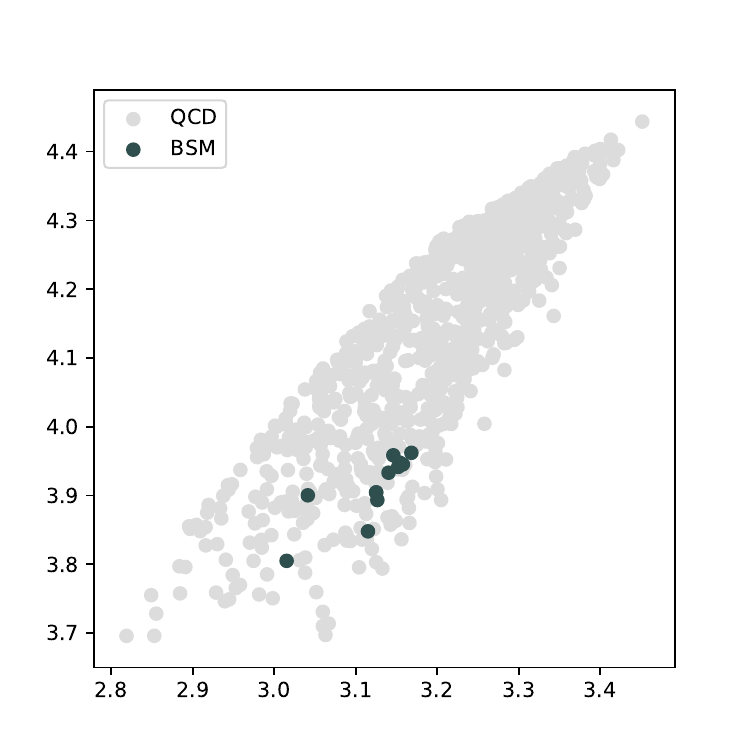}
    \includegraphics[width=0.40\textwidth]{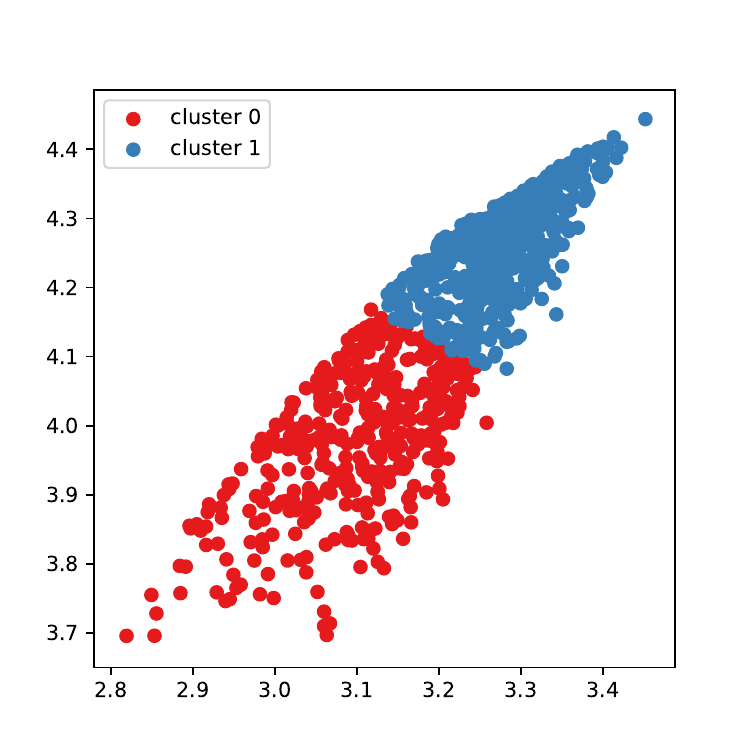}
    \caption{Visualization of the embedding space created for anomaly detection using 1000 events. Since the embedding space is already two-dimensional, no additional transformation is applied. The true labels are show on the left, while the clusters created by UCluster are shown on the right.}
    \label{fig:tsne_anomaly}
\end{figure*}

Here we would also like to point out that, even if a proxy of jet masses is given as an input, the trivial solution is still not achieved, since the model also has to identify which particles belong to which jets. To quantify the performance of UCluster, we start by considering only two clusters with an embedding space of same dimension. Figure \ref{fig:tsne_anomaly} shows the resulting embedding space without any transformation for 1000 random events.

\begin{table*}[ht]
    \centering
    \caption{Descriptions of each feature used to define a point in the point cloud implementation for anomaly detection. The last two lines are the global information added to parameterize the network.}
    \label{tab:anomaly_vars}
	\begin{tabular}{ll}

             Variable & Description  \\
            \hline
            $\Delta\eta$       &  \small{Pseudorapidity difference between the constituent and the associated jet}\\  
            $\Delta\phi$       &  \small{Azimuthal angle difference between the constituent and the  associated jet}\\  
            $\log\pt$       &  \small{Logarithm of the constituent's $\pt$ }\\  
            $\log\mathrm{E}$       &  \small{Logarithm of the constituent's E }\\  
            $\log\frac{\pt}{\pt(\mathrm{jet})}$       &  \small{Logarithm of the ratio between the constituent's $\pt$ and the associated jet $\pt$}\\  
            $\log\frac{\mathrm{E}}{\mathrm{E}(\mathrm{jet})}$       &  \small{Logarithm of the ratio between the constituent's E and the associated jet E}\\  
            $\Delta\mathrm{R}$       &  \small{Distance in the $\eta-\phi$ space between the constituent and the associated jet}\\  
            \hline
            $\log m_{J\{1,2\}}$ & \small{Logarithm of the masses of the two heaviest jets in the event} \\
            $\tau_{21}^{\{1,2\}}$ & \small{Ratio of $\tau_1$ to $\tau_2$ for the two heaviest jets in the event, with $\tau_N$ defined in \cite{Thaler2011}}  \\

	\end{tabular}
\end{table*}

\begin{figure*}[ht]
    \centering
    \includegraphics[width=0.40\textwidth]{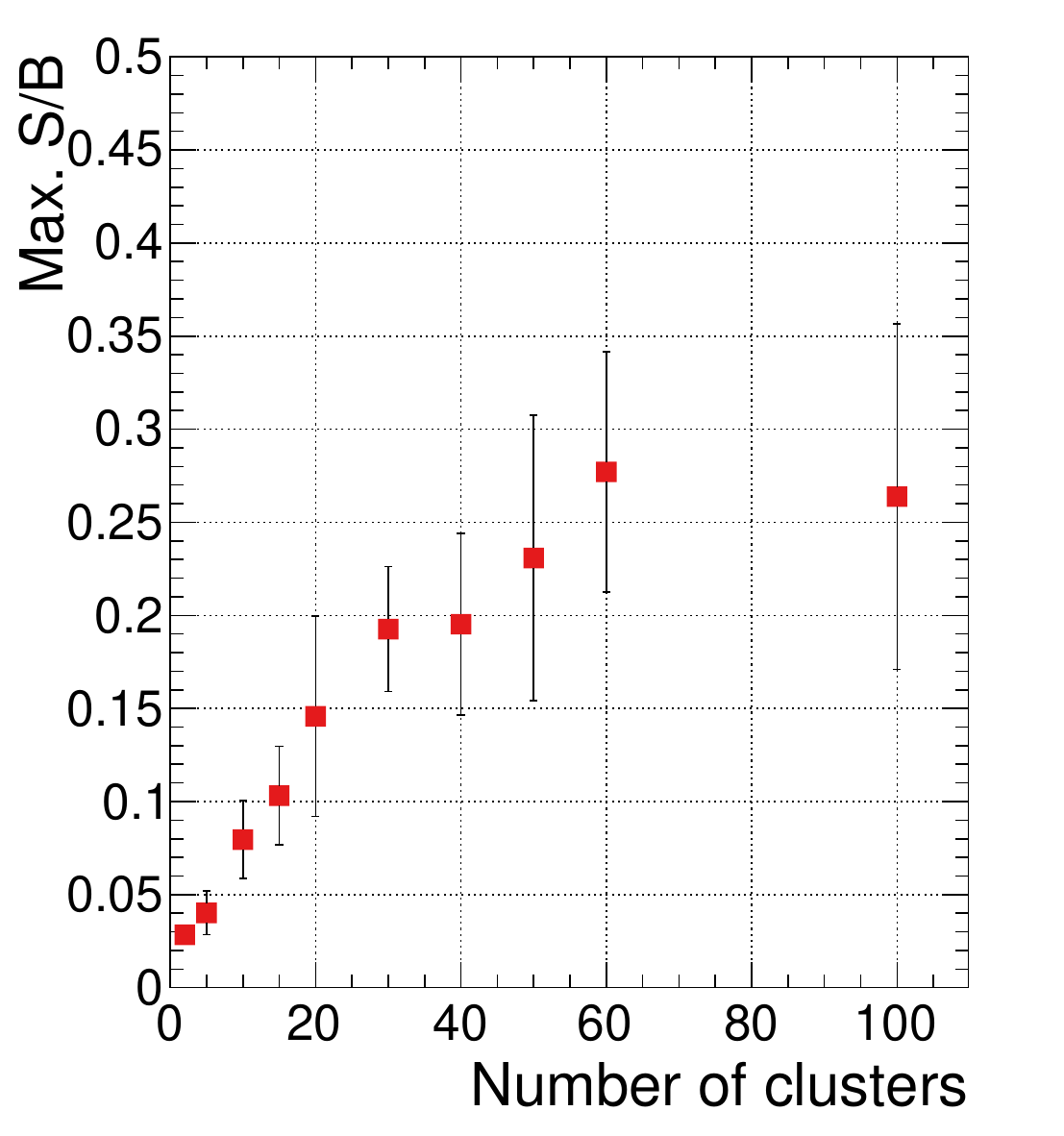}
    \includegraphics[width=0.40\textwidth]{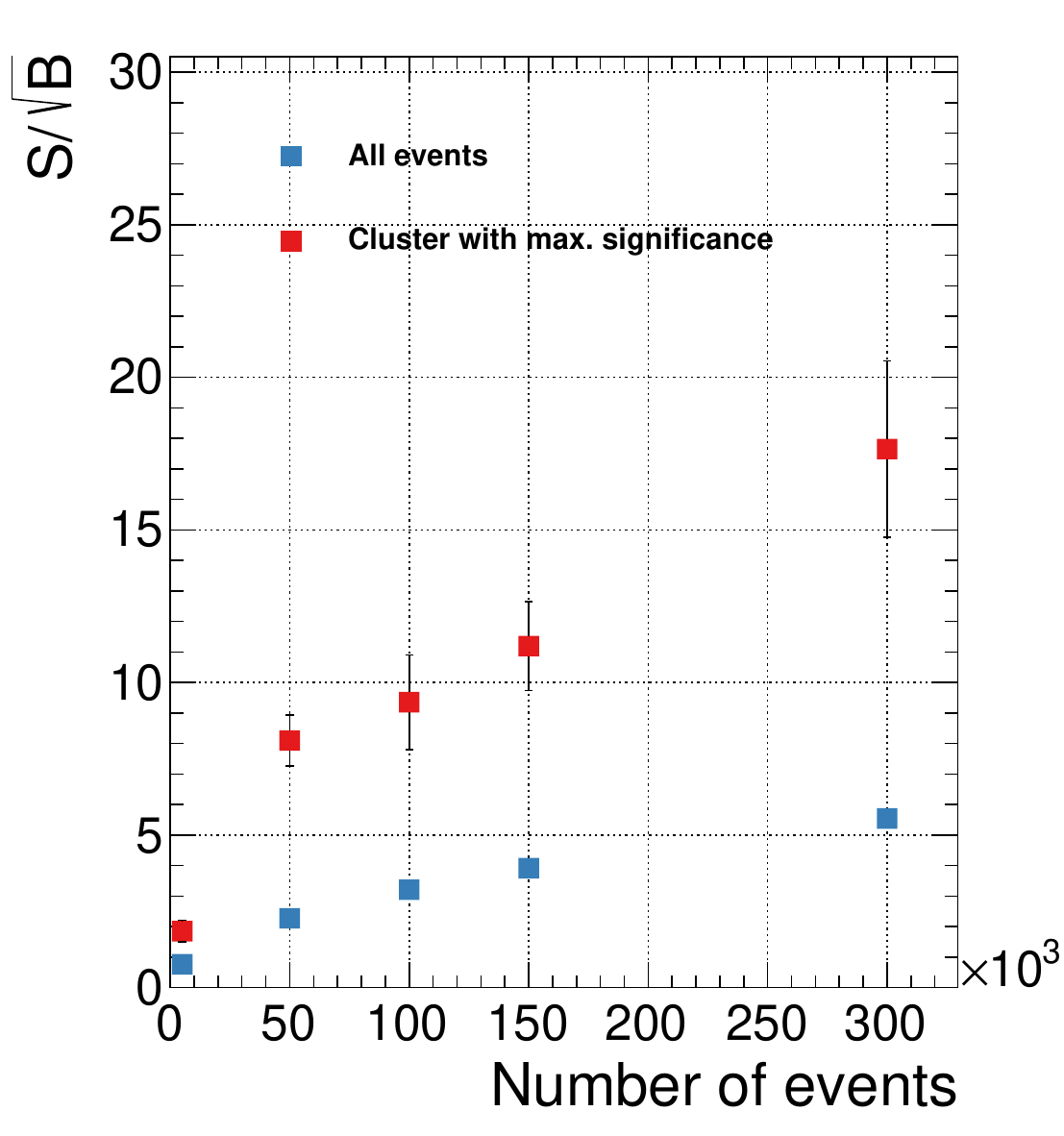}
    \caption{Maximum signal-to-background ratio found for different clustering sizes (left) and maximum significance found for UCluster trained and evaluated on different number of events with cluster size fixed to 30 (right). The uncertainty shows the standard deviation of the results from five trainings with different random weight initialization.}
    \label{fig:anomaly_sig}
\end{figure*}

Most of the BSM events are found in the same trained cluster, confirming the assumption that the signal events would end up close together in the embedding space. However, because of the large QCD background contamination present in the same cluster, the signal-to-background (S/B) ratio  remains low, increasing only from 1\% to 2.5\%. If the proximity assumption holds, then the cluster S/B ratio can be further enhanced by partitioning the events into more clusters. Indeed, if the classification loss favors an embedding space where signal events remain close together, increasing the number of clusters will decrease the QCD contamination in the signal clusters whose properties differ from the signal events. To test this assumption, the cluster size is varied while keeping all the other network parameters fixed. The maximum S/B ratio found in a cluster for different clusters sizes is shown in Fig.~\ref{fig:anomaly_sig} left. The S/B ratio steadily increases with cluster size, reaching an average of around 28\%. To test how the performance changes with the number of events, different training sample sizes were used while keeping the model fixed, the signal fraction fixed to 1\%,  and number of clusters fixed to 30. The result of each training is then evaluated in an independent sample which is the same size as the training sample. The result of the approximate significance (S/$\sqrt{\mathrm{B}}$) is shown in Fig.~\ref{fig:anomaly_sig} on the right. For initial significance in the range 2-6, we observe enhancements by factors 3-4.

The uncertainties in Fig.~\ref{fig:anomaly_sig} show the standard deviation of five independent trainings with different random initial weights. When many clusters are used, the clustering stability starts to decrease, as evidenced by the larger error bars. This behavior is expected, since a large cluster multiplicity requires clusters to target more specific event properties that might differ in between different trainings.

\begin{figure*}[ht]
    \centering
    \includegraphics[width=0.40\textwidth]{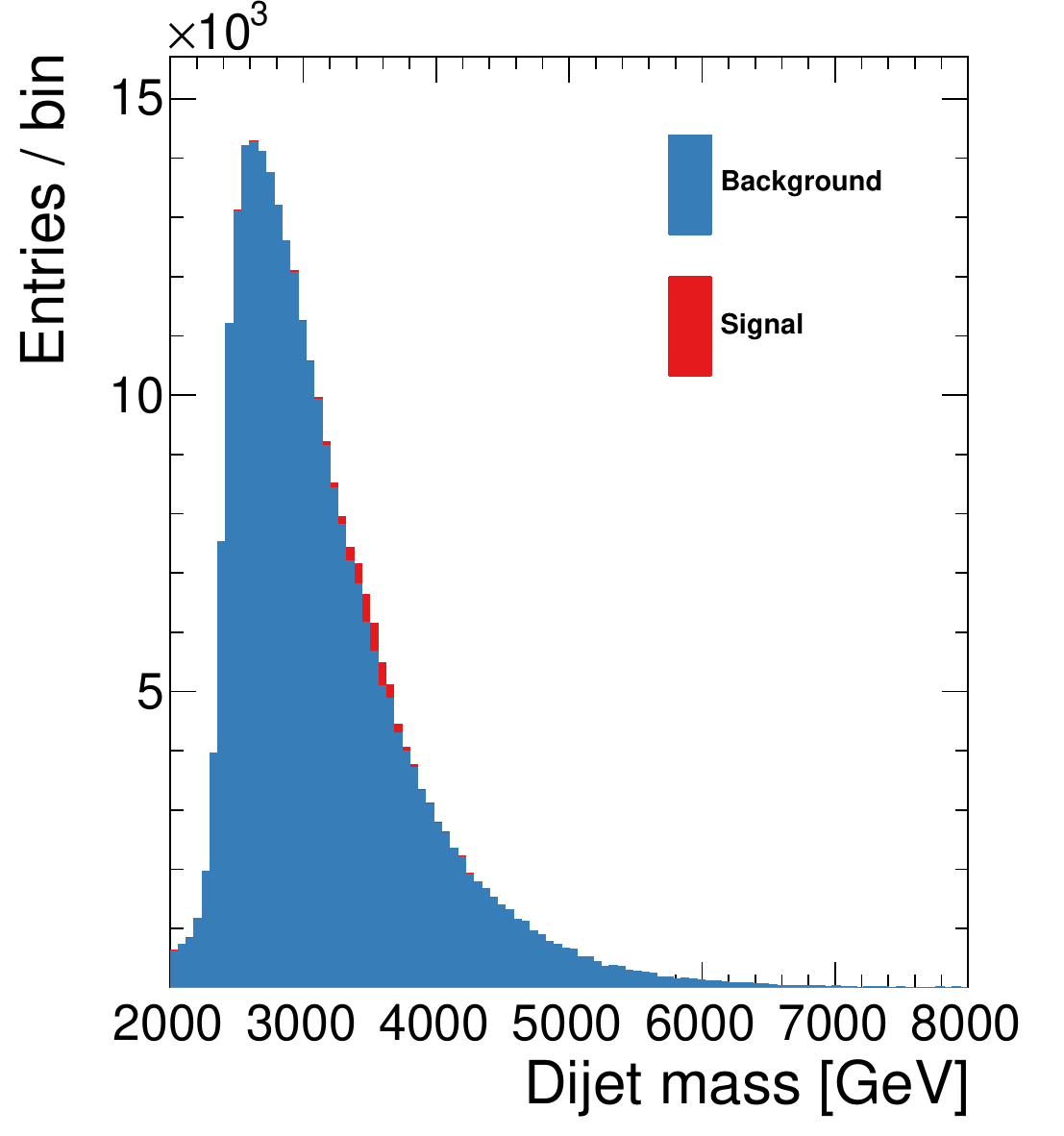}
    \includegraphics[width=0.40\textwidth]{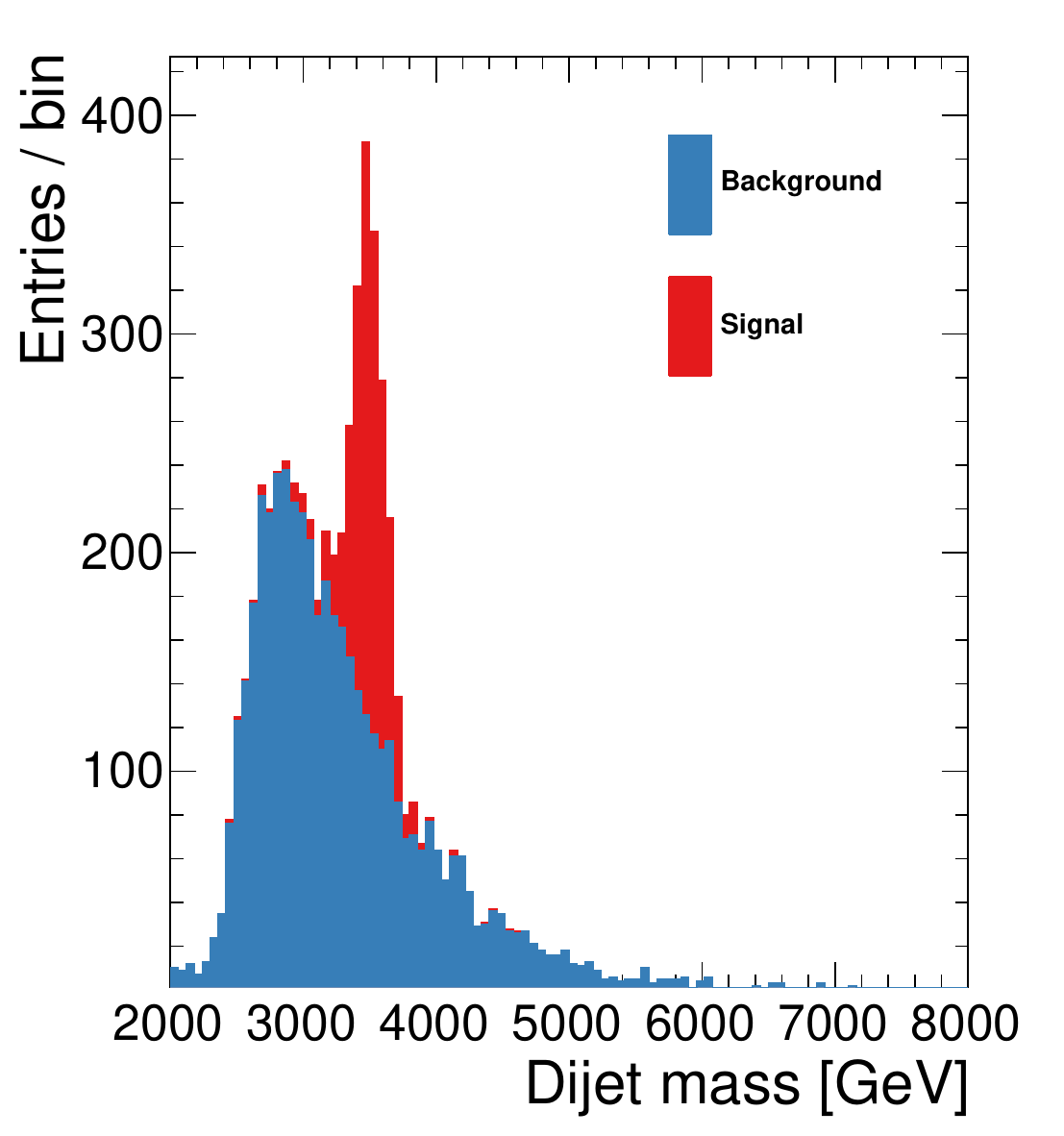}
    \caption{Dijet mass distribution of the events prior to clustering (left) and for the cluster with the highest S/B ratio (right), found when the data are partitioned into 60 clusters.}
    \label{fig:anomaly_mass}
\end{figure*}

To qualitatively verify the cluster composition, the dijet mass distribution for all events (left) and for the cluster with the highest S/B ratio (right) are shown in Fig.~\ref{fig:anomaly_mass}. 

\subsection{Background estimation}
In the previous section, the sensitivity to an anomalous signal was shown to improve with the number of clusters required by UCluster. However, requiring a larger number of clusters also requires a method to select interesting partitions for further inspection. A local p-value for each cluster can be determined for a background-only hypothesis, where the cluster with the lowest p-value is selected for further investigation. We also note that a global p-value can be derived by taking into account the  look-elsewhere effect \cite{Gross:2010qma}, which is already mitigated by the usage of independent samples during training, testing, and evaluation of UCluster. The main difficulty to estimate the p-value is to have a reliable background estimation for each cluster. Given that UCluster is encouraged to create clusters with more specific properties, the background shape for a given partition might not have a trivial description. A possible way to mitigate this issue is to use the nearest cluster, in embedding space, as a background model for the cluster under study. Given the anomalous signal remains localized in a particular cluster, the nearest cluster have the benefit to be signal free while still retaining similar properties to the cluster under consideration. To exemplify this idea, the  cluster with the highest S/B ratio shown in Fig.~\ref{fig:anomaly_mass} is used. To model the data distribution in the closest cluster, a smooth falling distribution with four free parameters, commonly used in dijet resonance searches is used \cite{Aaltonen:2008dn,ATLAS:2015nsi,Khachatryan:2016ecr}, described as:
\begin{equation}
    \frac{dN}{dm_{jj}} = p_0\frac{\left ( 1-x \right )^{p_1}}{x^{p_2 + p_3\ln(x)}}, x = m_{jj}/ \mathrm{1 TeV}.
\end{equation}

\begin{figure*}[ht]
    \centering
    \includegraphics[width=0.45\textwidth]{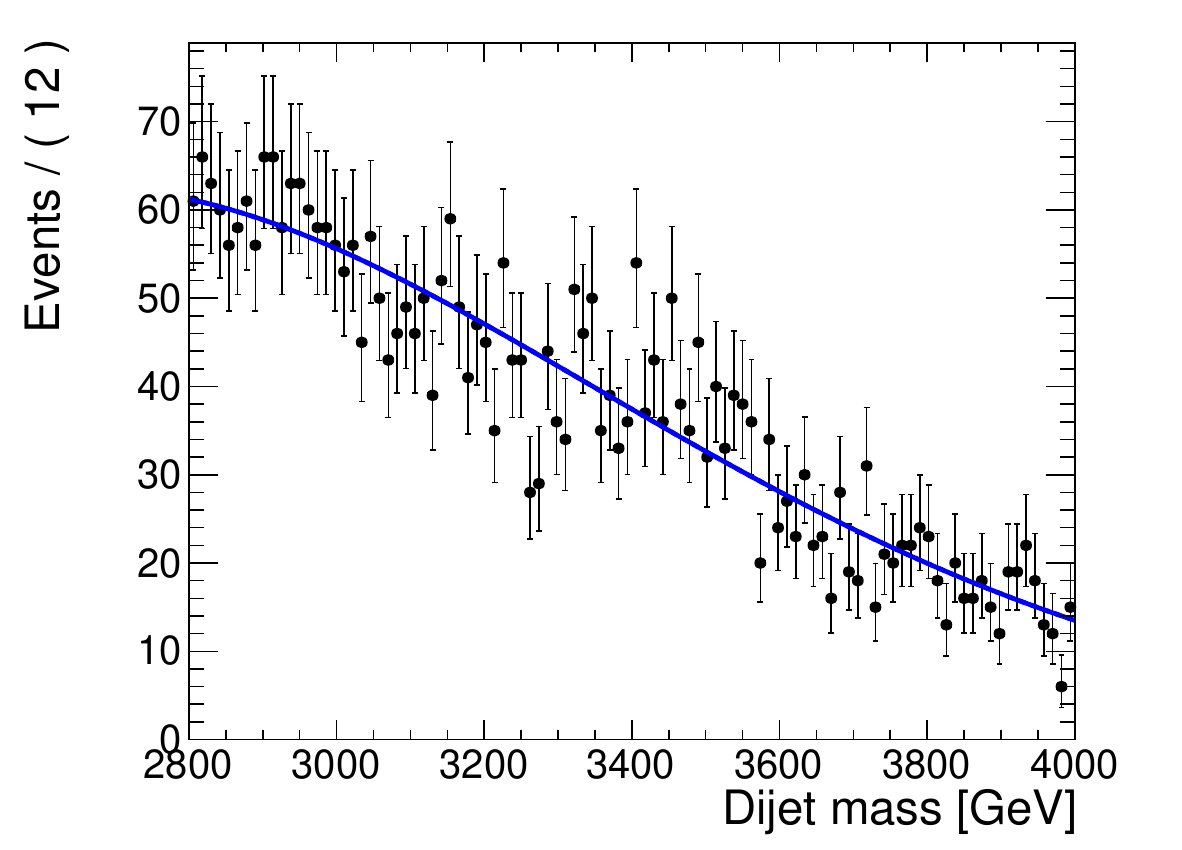}
    \includegraphics[width=0.45\textwidth]{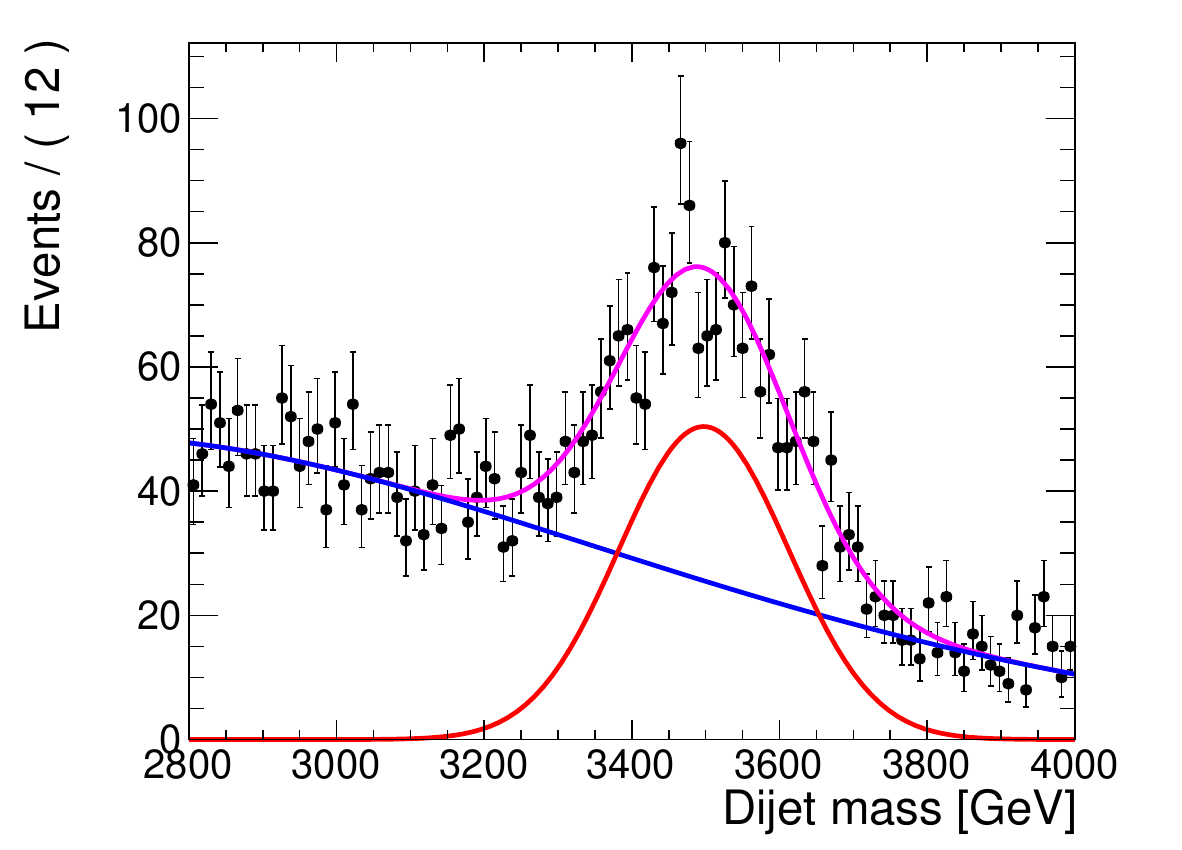}
    \caption{Dijet mass distribution for events in the cluster closest to the cluster with the highest S/B ratio (left) and the events in the cluster with the highest S/B ratio (right). The background component (blue) is determined in the closest cluster and extrapolated to the highest S/B ratio cluster. The signal contribution is shown in red while the sum of signal and background contributions are shown in magenta.}
    \label{fig:bkg_est}
\end{figure*}

After the fit is performed, all background parameters, besides the overall normalization, are kept frozen. This function is then used to model the background in the cluster with the highest S/B ratio. The signal modeling is done with a Gaussian function. The results of both fits are shown in Fig.~\ref{fig:bkg_est}. 

\subsection{Global distribution effects on clusters.}
In order to relate the clusters in embedding space to physical observables, four high-level features were added to the anomaly detection model: the invariant mass and $\tau_{21}$ of the two heaviest jets in the event.

\begin{figure*}[ht]
    \centering
    \includegraphics[width=0.40\textwidth]{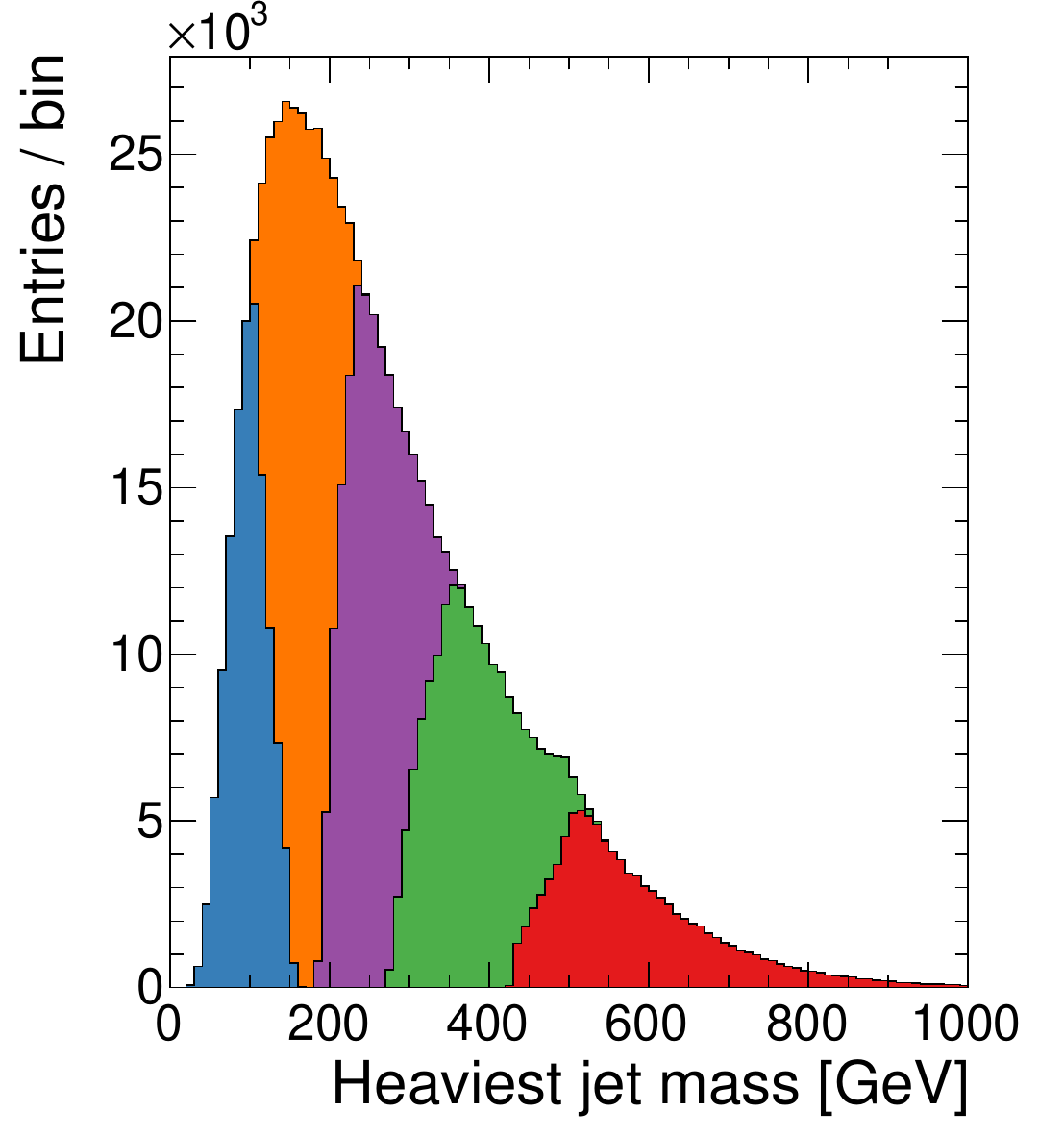}
    \includegraphics[width=0.40\textwidth]{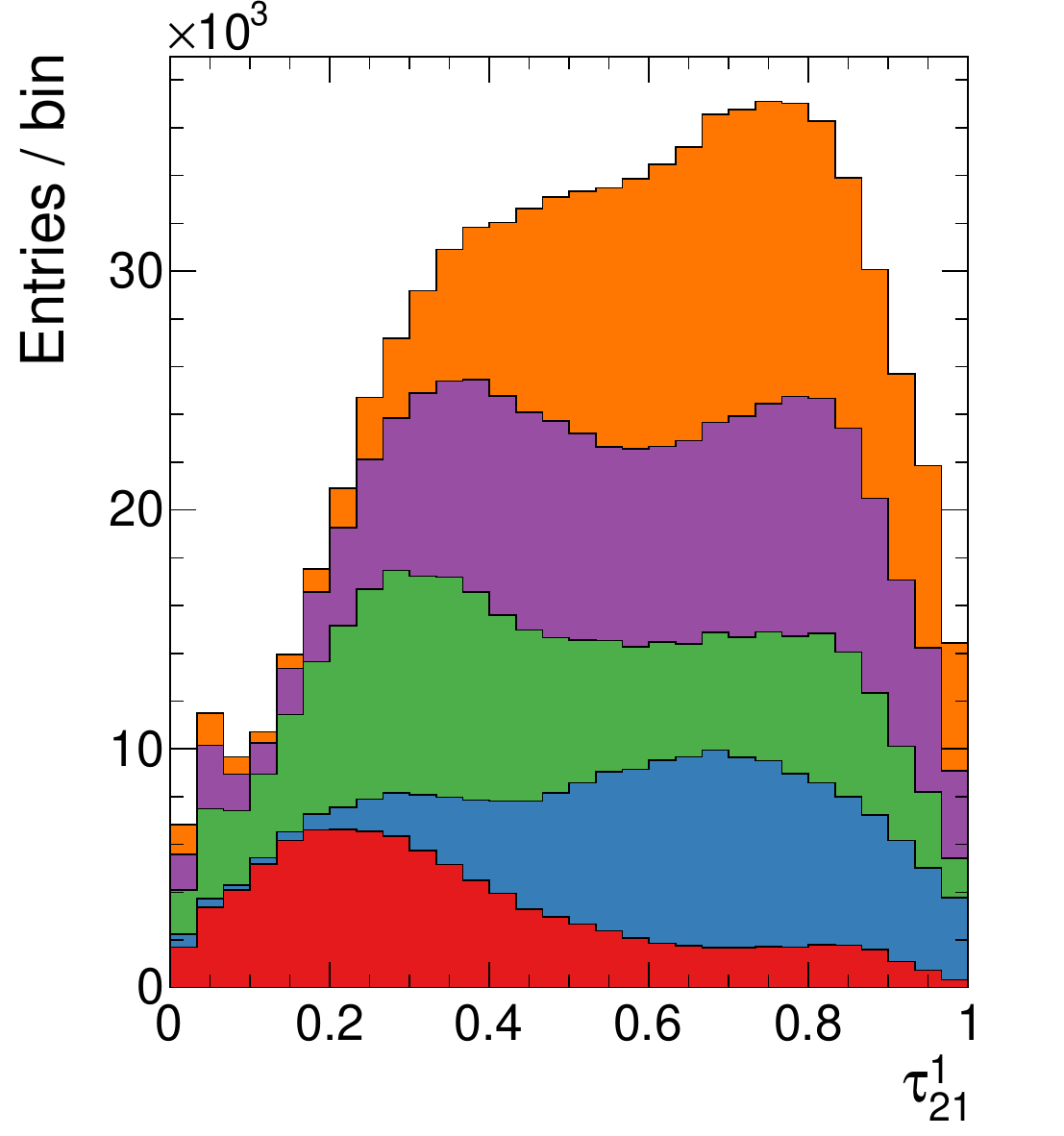}
    \includegraphics[width=0.40\textwidth]{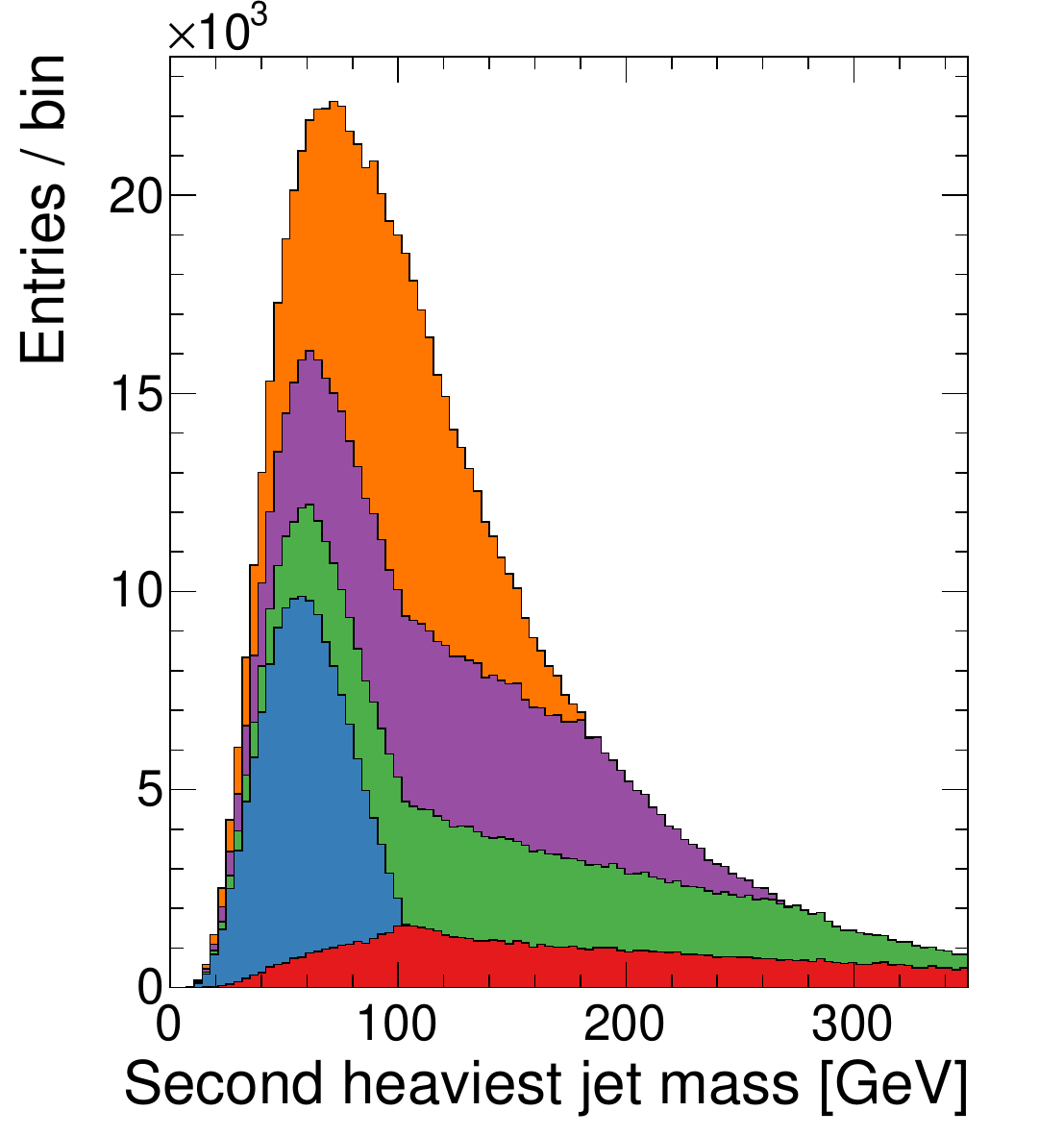}
    \includegraphics[width=0.40\textwidth]{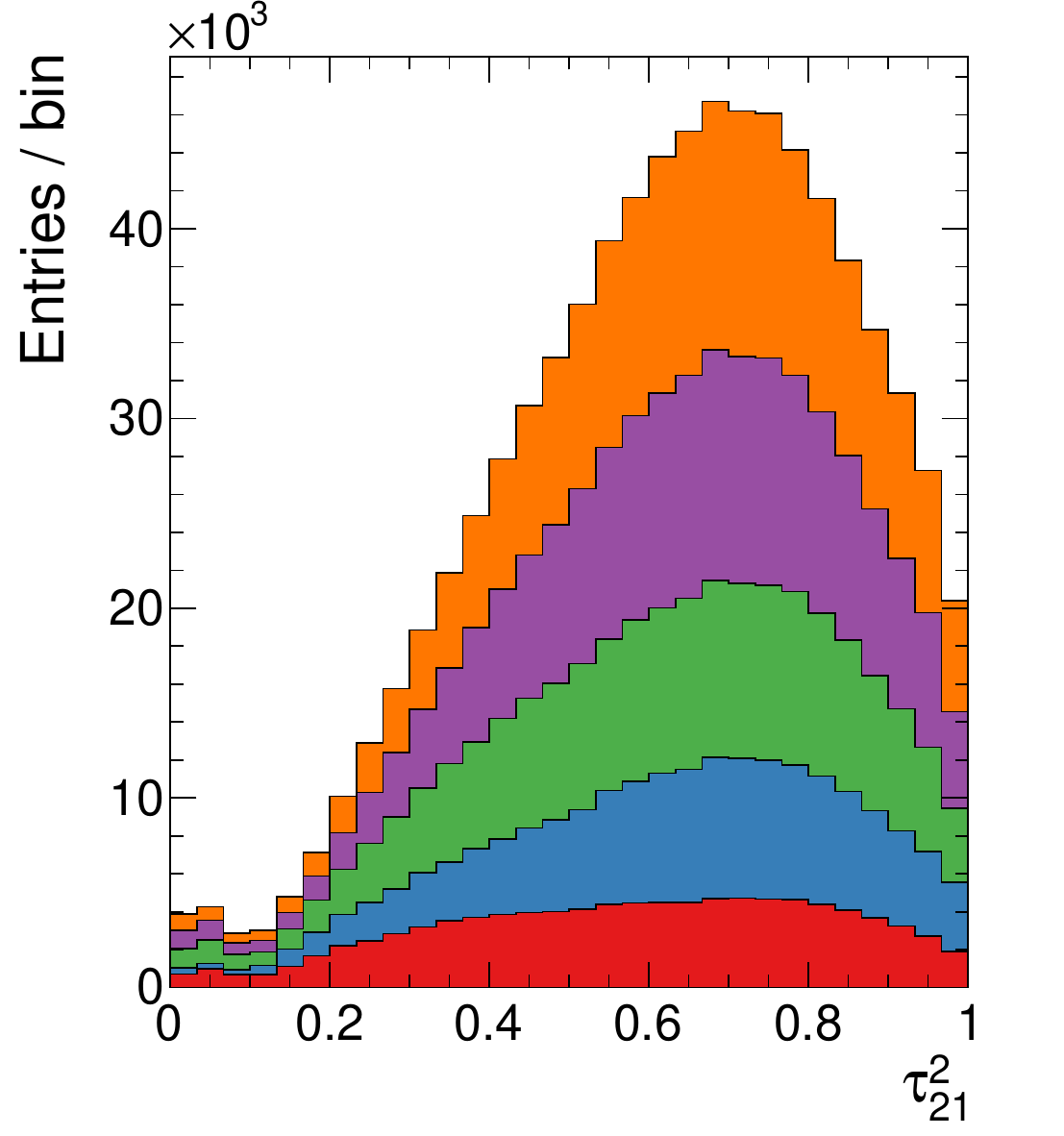}
    \caption{Distributions for the 4 high level features used to parameterize the performance of UCluster trained with 5 clusters. Events belonging to the same clusters receive the same color. The stacked contribution of all clusters is then shown.}
    \label{fig:stack_global}
\end{figure*}

To visualize the physical properties of the clusters, histograms of these four observables are shown in Fig.~\ref{fig:stack_global} with the stacked contributions of each individual cluster shown for UCluster with 5 clusters. From these distributions, there is a sharp separation between the cluster boundaries for the mass of heaviest jet in the dijet event. The sharp separation in jet mass is also related to the separation that is observed in the heaviest jet $\tau_{21}$. As pointed out in \cite{Komiske:2019fks}, QCD jets show a more distinctive two-prong structure when they have a larger mass. Therefore, heavier jets tend to have lower values of $\tau_{21}$. This correlation between jet mass and jet substructure is why the jet mass classification task leads to clusters where jets within a cluster have similar substructure.

\section{Conclusion and future prospects}

In this work, we presented UCluster, a new method to perform unsupervised clustering for collision events in high energy physics. We explored two potential applications for this method: unsupervised multiclass classification and anomaly detection. 

The ability of the embedding space to separate different processes is directly connected to the secondary task used in conjunction with the clustering objective. We proposed a classification task which was motivated by the observations of the correlation between the jet mass and jet substructure observables which is  often useful for jet tagging. By learning to classify the mass of a jet, UCluster created an embedding space that was shown to have a better separation power for all the class components in the data set compared to the jet mass alone.

UCluster was also studied for unsupervised anomaly detection. In this context, the classification task on jet masses was expanded to cover the entire event topology. Using this method, we were able to increase the signal-to-background ratio in a given cluster from an initial value of 1\% up to 28\%, while also observing a stable performance even for a large cluster multiplicity. A data-driven background estimation is also possible by using the closest cluster in embedding space to the cluster under investigation. This data-driven method allows the selection of interesting clusters by comparing the background compatibility with the nearest cluster. Clusters of interest can be further investigated by a dedicated analysis.

We remark that different tasks than the ones proposed in this work can also be used to create meaningful embeddings. In particular, recent advances in auto-encoders applied to particle physics \cite{ATL-SOFT-PUB-2018-001} are strong candidates for a summary statistic that can encapsulate the event information in a lower dimensional representation, suitable for clustering.

Compared to \cite{JMLR:v10:quadrianto09a,NIPS2014_5453}, we relax the requirements on the label proportion for each different component in a mixed sample. One interesting point to notice is that, as presented in \cite{Genevay2019DifferentiableDC}, the clustering assignment problem can instead be interpreted as an optimal transport problem. This insight is particularly interesting when the label proportions are known a priori. In this case, the additional knowledge of the label proportions can be directly added to the model as a regularization term of the form: 

\begin{eqnarray}
    L_{\mathrm{reg. cluster}} &&= min \sum_k ^K \sum_j^n \left \| f_{\theta}(x_j) - \mu_k \right \|^{2}\pi_{jk} \nonumber\\
    && + \alpha \pi_{jk} (\log(\pi_{jk}) - 1).
\end{eqnarray}

This approach requires the term $\pi_{jk}$ to be numerically solved, subject to:
\begin{eqnarray}
     &&\pi\bold{1}_K = \frac{1}{n}\bold{1}_N, \nonumber \\
     &&\pi^\mathrm{T}\mathbf{1}_N = w,
\end{eqnarray}
where $w$ represent the vector of label proportions.

Furthermore, we considered an application where the initial number of mixed components was known. This condition was necessary to select a suitable number of clusters. However, this requirement could also be relaxed, as shown in  \cite{ren2018deep,Patil_2019}, for example, where the clustering model is able to identify the optimal number of partitions given the properties of a data set.

Finally, UCluster can also be used in conjunction with other anomaly detection approaches, where first a set of interesting clusters are identified and then further inspected by other methods.

\begin{acknowledgments}
The authors would like to thank Kyle James Read Cormier for the valuable suggestions regarding the development and clarity of this document. This research was supported in part by the Swiss National Science Foundation (SNF) under contract No. 200020-182037.
\end{acknowledgments}

\bibliography{apssamp}% Produces the bibliography via BibTeX.

\end{document}